\documentclass[11pt]{article}
\usepackage{moriond,epsfig}
\usepackage{floatflt,graphicx}
\usepackage{wrapfig}

\def\NIMA{{\em Nucl. Instrum. Methods} A~}
\def\NPB{{\em Nucl. Phys.} B~}

\def\PRL{{\em Phys. Rev. Lett.}~}

\def\cpc{{\em Comp. Phys. Comm}~}
\def\epj{{\em Eur Phys J.}~}
\def\np{{\em Nucl. Phys.}~}

\def\be{\begin{equation}}
\def\ee{\end{equation}}
\def\bea{\begin{eqnarray}}
\def\eea{\end{eqnarray}}

\begin{document}
\vspace*{4cm}
\title{QCD dynamics at low $x_{\rm Bj}$ in $ep$ collisions at HERA}

\author{ T.E. DANIELSON~\footnote{On behalf of the H1 and ZEUS collaborations} }

\address{Department of Physics, University of Wisconsin - Madison, 1150 University Avenue,\\
Madison, WI 53706, USA}

\maketitle\abstracts{Forward jet and multijet production has been measured at low Bjorken
  $x$ at HERA.  The measured cross sections and correlations were compared to predictions from 
  DGLAP-based fixed-order calculations.  Further comparisons were made to DGLAP-based and CCFM-based 
  leading-order Monte Carlo predictions, as well as to Colour-Dipole model predictions.  For the majority of
  the phase space covered in the HERA kinematic region, fixed-order calculations describe the data well, while
  the leading-order models provide an inconsistent description of the data.}

\section{Introduction}

Jet production in DIS is an ideal environment for
investigating different approaches to parton dynamics at low Bjorken-$x$, $x_{\rm Bj}$.
An understanding of this regime is of particular
relevance in view of the startup of the LHC, where many of the Standard Model processes such as the production
of electroweak gauge bosons or the Higgs particle involve the collision of partons with
a low fraction of the proton momentum.
 
In the usual collinear QCD factorisation approach, the cross sections are obtained as the convolution of
perturbative matrix elements and parton densities evolved according to the
DGLAP evolution equations.  
In these equations, all orders proportional to $\alpha_s \ln Q^2$ and terms with double logarithms $\ln Q^2 \cdot \ln{1/x}$,
where $x$ is the fraction of the proton momentum carried by a parton, which is equal to $x_{\rm Bj}$ in the
quark-parton model, are resummed.  In the DGLAP approach, the parton participating 
in the hard scattering is
the result of a partonic cascade ordered in transverse momentum, $p_T$.  The partonic cascade starts from a low-$p_T$
and high-$x$ parton from the incoming proton and ends up, after consecutive branching, in the high-$p_T$
and low-$x$ parton entering in the hard scattering.
At low $x_{\rm Bj}$, where the phase space for parton emissions increases, terms proportional to
$\alpha_s \ln 1/x$ may become large and spoil the accuracy of the DGLAP approach.  In this phase-space region, 
a better description may come from the BFKL approach, which resums terms proportional to 
$\ln 1/x$, and the CCFM approach, which uses unintegrated gluon densities in an all-loop non-Sudakov resummation.  

Parton evolution schemes at low $x_{\rm Bj}$ were studied at HERA by measuring forward jet production and 
correlations in jet angles and transverse momentum.  An excess of forward jets compared to DGLAP-based predictions 
and jets produced in the hard scatter that are not strongly correlated in transverse momentum may indicate the 
breakdown of DGLAP dynamics.

\section{Dijet Azimuthal Correlations}

\begin{wrapfigure}{r}{7.0cm}
  \centering\vspace{-0.5cm}
  \includegraphics[width=6.5cm]{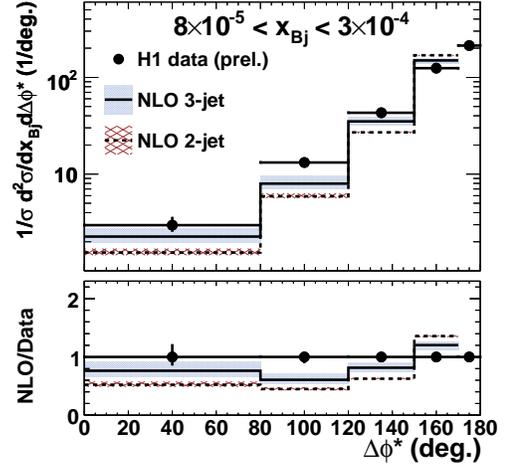}    
  \caption{
    Double-differential normalised (see text) cross sections as a function of $\Delta\phi^{*}$ as measured by H1 
    compared to {\sc NLOjet} calculations for $\mathcal{O}(\alpha_{s}^2)$ and $\mathcal{O}(\alpha_{s}^3)$.
    } 
  \label{fig_azim_corr_NLO}
\end {wrapfigure}

Dijet azimuthal correlations were investigated by the H1 Collaboration\cite{H1-prelim-06-032:2006} by measuring the 
cross-sections $d^{2}\sigma/dx_{\rm Bj}d\Delta\phi^{*}$, where $\Delta\phi^{*}$ is the azimuthal separation in the 
hadronic centre-of-mass (HCM) frame between the two selected jets closest to the scattered electron in pseudorapidity, 
$\eta$.  The measurements of $\Delta\phi^{*}$ are reasonably well-described by {\sc NLOjet} \cite{prl:87:082001} 
calculations at $\mathcal{O}(\alpha_{s}^3)$, 
albeit within large theoretical uncertainties.  To reduce the theoretical uncertainties, the measurements were normalised to the 
visible cross section for $\Delta\phi^{*} < 170^{\circ}$.  With a reduced theoretical uncertainty, the calculations are 
shown to predict a narrower $\Delta\phi^{*}$ spectrum than is measured, especially at very low $x_{\rm Bj}$, as shown
in Fig.~\ref{fig_azim_corr_NLO}.  The measurements were also compared to predictions from two {\sc Rapgap} \cite{cpc:86:147} 
(DGLAP) samples, with one sample using only direct photons, the other using both direct and resolved photons; 
{\sc Lepto} \cite{cpc:101:135} (CDM); and {\sc Cascade} \cite{epj:c19:351} (CCFM).  All models fail to describe $\Delta\phi^{*}$ 
over the entire range in $x_{\rm Bj}$ covered.

\section{Multijet Correlations}

\begin{wrapfigure}{r}{7.0cm}
  \centering\vspace{-1.0cm}
  \includegraphics[width=6.5cm]{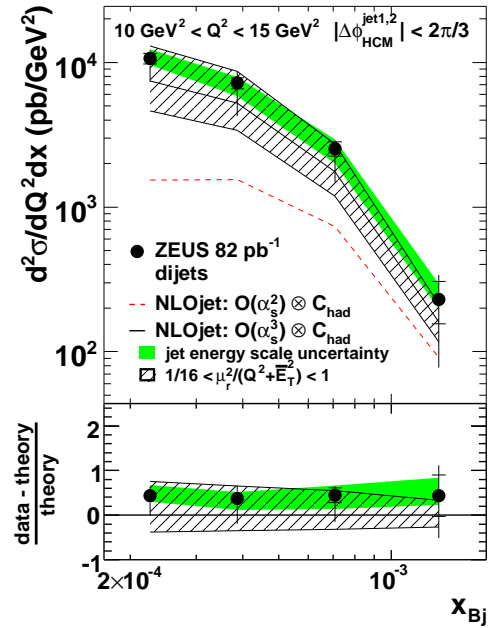}
  
  \caption{ZEUS dijet cross sections in $x_{\rm Bj}$ and $Q^{2}$ as a function of $x_{\rm Bj}$ with 
    $\Delta\phi_{HCM}^{\rm jet 1,2} < 120^{\circ}$}
  
  \label{fig_low_delphi}
\end{wrapfigure}

The sensitiviy of parton evolution to the topology of the jet system was studied by the ZEUS 
collabortion \cite{Collaboration:2007dx}. 
Multi-differential cross sections as functions of the jet correlations in transverse momenta, azimuthal angles, 
and pseudorapidity have been measured for dijet and trijet production in the HCM frame.  DGLAP-based 
calculations from {\sc NLOjet} at $\mathcal{O}(\alpha_{s}^2)$ 
and $\mathcal{O}(\alpha_{s}^3)$ were compared to the measurements.  The {\sc NLOjet} calculations at $\mathcal{O}(\alpha_{s}^2)$ 
do not describe the correlations in transverse momenta and azimuthal angle for dijet events; however with inclusion of 
higher-order terms, the {\sc NLOjet} calculations at $\mathcal{O}(\alpha_{s}^3)$ describe the dijet data over the 
entire range in $x_{\rm Bj}$ covered.  The importance of higher-order terms at low $x_{\rm Bj}$ is seen especially when measuring 
the double-differential cross sections in $Q^2$ and $x_{\rm Bj}$ for events with $\Delta\phi_{HCM}^{\rm jet 1,2} < 120^{\circ}$, 
where $\Delta\phi_{HCM}^{\rm jet 1,2}$ is the azimuthal separation of the two jets with the highest transverse energy.  At low
$x_{\rm Bj}$, the {\sc NLOjet} calculations at $\mathcal{O}(\alpha_S^{3})$ are up to about one order of magnitude larger 
than the $\mathcal{O}(\alpha_S^{2})$ calculations and are consistent with the data, as seen presented in Fig.~\ref{fig_low_delphi}.  
The {\sc NLOjet} calculations at $\mathcal{O}(\alpha_{s}^3)$ also provide a reasonable description of the trijet measurements, 
with the description improving somewhat at higher $x_{\rm Bj}$.

\section{Forward Jet Production}

\begin{figure}[htp]
  
  \begin{center}
    
    \begin{minipage}{0.83\linewidth}    
      
      \epsfig{figure=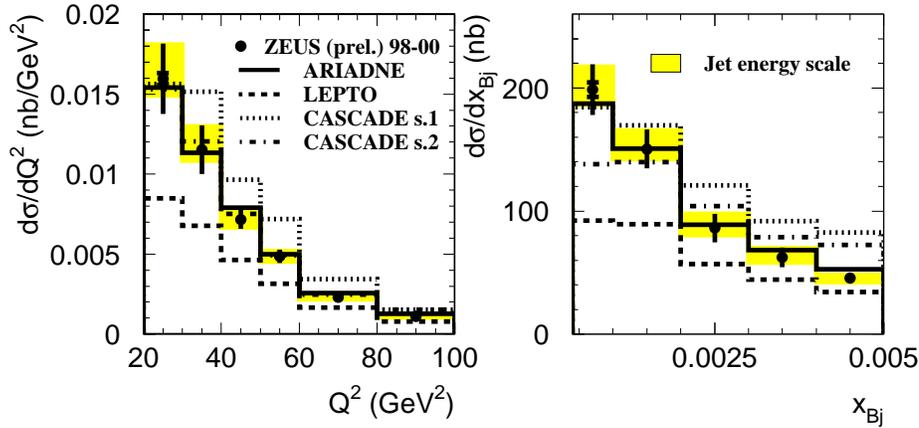,width=\linewidth}
      
      \caption{ZEUS forward jets as a function of the kinematic variables $x_{\rm Bj}$ and $Q^{2}$ compared to predictions from
        {\sc Ariadne}, {\sc Lepto}, and {\sc Cascade}}
      
      \label{zeus_forward_fig}
    \end{minipage}
  \end{center}
\end{figure}

To examine the sensitivity of parton evolution to forward jet production, the ZEUS collaboration has studied jet production
in an extended pseudorapidity range of $\eta_{LAB}^{\rm jet} < 3.5$ by incorporating the Forward Plug Calorimeter (FPC)
 \cite{nim:a450:235} used during the HERAI running period \cite{ZEUS-prel-05-014:2005}.  
Meaurements of cross sections as functions of $Q^2$, $x_{\rm Bj}$, 
$E_{T,LAB}^{\rm jet}$, and $\eta_{LAB}^{\rm jet}$ are reasonably well-described by DGLAP-based calculations from 
{\sc Disent} \cite{np:b485:291}, with large 
theoretical uncertainties at both low $x_{\rm Bj}$ and high $\eta_{LAB}^{\rm jet}$.  Predictions from {\sc Lepto} 
(DGLAP); {\sc Ariadne} \cite{cpc:71:15} (CDM); and {\sc Cascade}, 
with two sets from the J2003 unintegrated gluon PDF used, were also compared to the measurements.  Overall, {\sc Ariadne} 
provides the best description of the measured cross sections; {\sc Lepto} consistently underestimates the cross sections, 
and {\sc Cascade} fails to consistently reproduce the shapes of the distributions (see Fig.~\ref{zeus_forward_fig}).

\section{Trijet Production and Correlations}

\begin{wrapfigure}{r}{7.0cm}
  \centering\vspace{-1.2cm}
  \includegraphics[width=6.5cm]{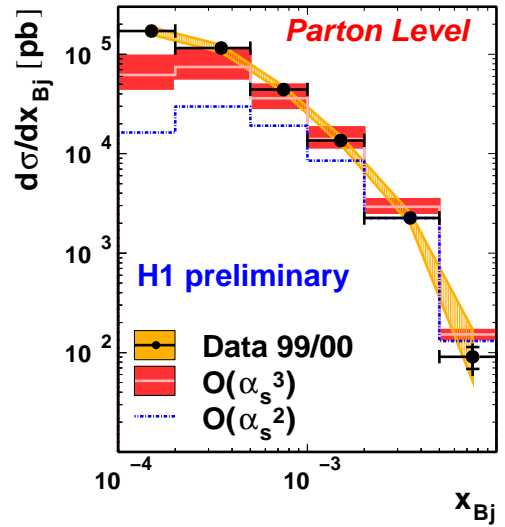}
  \caption{H1 trijet cross sections as a function of $x_{\rm Bj}$ for events with two forward jets compared
    to {\sc NLOjet} calculations at $\mathcal{O}(\alpha_{s}^2)$ and $\mathcal{O}(\alpha_{s}^3)$.}
  \label{H1_trijet_forward}
\end{wrapfigure}

Trijet cross sections and correlations were measured by the H1 collaboration as a study of parton evolution at low 
$x_{\rm Bj}$ \cite{H1-prelim-06-034:2006}. 
Cross sections were measured as functions of $x_{\rm Bj}$, jet pseudorapidity, scaled jet energies, and correlations 
in the jet angles $\theta^{'}$ and $\psi^{'}$.  The variable $\theta^{'}$ is defined as the angle between 
the proton beam and the jet with the highest transverse energy, while $\psi^{'}$ is defined as the angle between 
the plane defined by the proton beam and the highest $E_{T}$ jet, and the plane defined by the two jets with the 
highest $E_{T}$.  These measurements were made for three separate trijet samples: an inclusive trijet sample, and 
two trijet samples with one and two forward jets, respectively, with a forward jet having 
$\theta_{LAB}^{\rm jet} < 20^{\circ}$ and $x_{jet} = E_{HCM}^{\rm jet}/E_{pbeam} > 0.035$.  For the inclusive trijet 
sample, {\sc NLOjet} calculations provide a reasonable description of the measured cross secitons, but slightly undersestimate 
the measurements in the lowest bin of $x_{\rm Bj}$.  The agreement between the calculations and the measured cross section in
$x_{\rm Bj}$ is worse for the trijet sample containing two forward jets, with the most noticable disgreement observed at lowest 
$x_{\rm Bj}$ (see Fig.~\ref{H1_trijet_forward}).  The selection of two forward jets favors events with 
forward gluon emission unorderd in transverse momentum, which the calculations at $\mathcal{O}(\alpha_{s}^3)$ do not predict 
entirely.  Also seen in Fig.~\ref{H1_trijet_forward} is that the higher-order terms in the {\sc NLOjet} calculations are important 
for forward jet emissions.   The other cross sections for this sample are well-described by the calculations. 

Predictions from {\sc Djangoh} (CDM) and {\sc Rapgap} LO MC models were also compared to the measured cross sections.  
The cross sections for the inclusive trijet sample are better described by CDM predictions, but both the 
CDM and {\sc Rapgap} predictions are inconsistent for the jet correlation angles $\theta^{'}$ and $\psi^{'}$; the 
{\sc Rapgap} predictions fail to describe the $\theta^{'}$ distributions, and the CDM predictions fail to 
describe the $\psi^{'}$ cross sections.

\section{Summary}

Parton dynamics at low $x_{\rm Bj}$ ($10^{-4} < x_{\rm Bj} < 10^{-2}$) have been investigated at HERA by the ZEUS 
and H1 collaborations.  DGLAP-based NLO calculations describe the measured cross sections and jet correlations reasonably well 
for the most part when higher-order terms in the calculations are properly taken into account.  The calculations fail to 
describe the trijet cross section in $x_{\rm Bj}$ when the trijet sample contains two forward jets.  
Leading-order Monte Carlo models provide an inconsistent description of the measured cross sections.  DGLAP-based LO MC 
models in general do not describe the cross sections; CDM models fail to describe $\Delta \phi^{*}$ and $\psi^{'}$; 
{\sc Cascade} predictions are highly sensitive to the unintegrated gluon PDF used, and do not describe the data consistently.

\end{document}